\documentclass[sigconf]{acmart}

\AtBeginDocument{%
  \providecommand\BibTeX{{%
    \normalfont B\kern-0.5em{\scshape i\kern-0.25em b}\kern-0.8em\TeX}}}

\setcopyright{none}

\renewcommand\footnotetextcopyrightpermission[1]{} 
\usepackage{booktabs}   
\usepackage{subcaption} 

\usepackage{xcolor}
\usepackage{soul}
\usepackage{xspace}
\usepackage[epsilon]{backnaur}
\usepackage{framed}
\usepackage{listings}
\usepackage{balance}
\usepackage{amsmath,amsfonts,amsbsy,bm,stmaryrd}
\usepackage[mathscr]{eucal}
\usepackage{xparse}
\usepackage{algorithmic}
\usepackage{graphicx}
\usepackage{wrapfig}
\usepackage{algorithm} 
\usepackage{wrapfig}
\usepackage{xargs}
\usepackage{subcaption}
\usepackage{cleveref}
\usepackage[acronym]{glossaries}
\makeglossaries

\usepackage{nicematrix}

\usepackage{tikz}
\usetikzlibrary{matrix}
\usepackage{multirow}
\usepackage{tabularx}
\makeatletter
\def\hlinewd#1{%
\noalign{\ifnum0=`}\fi\color{white}\hrule\@height #1 %
\futurelet\reserved@a\@xhline}
\makeatother

\newlength{\Oldarrayrulewidth}

\NewDocumentCommand \T { O{} m } {\ensuremath{\boldsymbol{#1\mathscr{\MakeUppercase{#2}}}}}
\NewDocumentCommand \M { O{} m } {\ensuremath{\bm{#1\mathbf{\MakeUppercase{#2}}}}} 
\NewDocumentCommand \V { O{} m } {\ensuremath{\bm{#1\mathbf{\MakeLowercase{#2}}}}} 

\definecolor{gokcen}{RGB}{255,50,255}
\definecolor{erdal}{rgb}{0.0,0.5,0.0}
\definecolor{roby}{RGB}{23,125,54}
\definecolor{jacques}{RGB}{50,125,154}
\definecolor{bin}{RGB}{255,0,0}

\ifx\paperversiondraft\undefined
\newcommand{\jp}[2][1=]{}
\newcommand{\gk}[2][1=]{}
\newcommand{\rg}[2][1=]{}
\newcommand{\emm}[2][1=]{}
\newcommand{\br}[2][1=]{}
\else
\newcommandx{\jp}[2][1=]{#1\textcolor{jacques}{\textbf{Jacques:} #2}}
\newcommandx{\gk}[2][1=]{#1\textcolor{gokcen}{\textbf{Gokcen:} #2}}
\newcommandx{\rg}[2][1=]{#1\textcolor{roby}{\textbf{Roberto:} #2}}
\newcommandx{\emm}[2][1=]{#1\textcolor{erdal}{\textbf{Erdal:} #2}}
\newcommandx{\br}[2][1=]{#1\textcolor{bin}{\textbf{Bin:} #2}}
\fi

\usepackage{mips}
\usepackage{color}

\definecolor{dkgreen}{rgb}{0,0.6,0}
\definecolor{gray}{rgb}{0.5,0.5,0.5}
\definecolor{mauve}{rgb}{0.58,0,0.82}

\definecolor{mygreen}{rgb}{0,0.6,0}
\makeglossaries

\usepackage[most]{tcolorbox}
\usepackage{colortbl}
\usepackage{hhline}
\usepackage{array}

\usepackage{tikz}
\usepackage{pgfplots}
\usetikzlibrary{positioning,arrows}
\usetikzlibrary{decorations.pathreplacing,calc}

\begin{document}


\title{COMET: A Domain-Specific Compilation of High-Performance Computational Chemistry}

\newcommand{\name}{COMET\xspace}%
\newcommand{\naive}{na\"{\i}ve\xspace}%

\title{COMET: A Domain-Specific Compilation of High-Performance Computational Chemistry}
%
%

\author{Erdal Mutlu, Ruiqin Tian}
\affiliation{%
 \institution{Pacific Northwest National Laboratory}
 \city{Richland, WA}
 \country{USA}}
\email{{erdal.mutlu, ruiqin.tian}@pnnl.gov}

\author{Bin Ren}
\affiliation{%
 \institution{William \& Mary}
 \city{Williamsburg, VA}
 \country{USA}}
\email{bren@cs.wm.edu}

\author{Sriram Krishnamoorthy}
\affiliation{%
 \institution{Pacific Northwest National Laboratory}
 \city{Richland, WA}
 \country{USA}}
\email{{sriram}@pnnl.gov}

\author{Roberto Gioiosa}
\affiliation{%
 \institution{Pacific Northwest National Laboratory}
 \city{Richland, WA}
 \country{USA}}
\email{{roberto.gioiosa}@pnnl.gov}

\author{Jacques Pienaar}
\affiliation{%
 \institution{Google}
 \city{Mountain View, CA}
 \country{USA}}
\email{jpienaar@google.com}

\author{Gokcen Kestor}\affiliation{%
 \institution{Pacific Northwest National Laboratory}
 \city{Richland, WA}
 \country{USA}}
\email{gokcen.kestor@pnnl.gov}

%
\renewcommand{\shortauthors}{E. Mutlu et al.}
\renewcommand{\shorttitle}{COMET: A Domain-Specific Compilation for Computational Chemistry}
%

\begin{abstract}
The computational power increases over the past decades have greatly enhanced the ability to simulate
chemical reactions and understand ever more complex transformations. Tensor contractions are the
fundamental computational building block of these simulations.
These simulations have often
been tied to one platform and restricted in generality by the interface provided to the user. The
expanding prevalence of accelerators and researcher demands necessitate a more general approach
which is not tied to specific hardware or requires contortion of 
algorithms to specific
hardware platforms. In this paper we present \name, a domain-specific programming language and
compiler infrastructure for tensor contractions targeting 
heterogeneous accelerators.
We present a system of progressive lowering through multiple layers of abstraction and optimization
that 
achieves up to $1.98\times$ speedup for 30 tensor contractions 
commonly used
in computational chemistry and beyond.

\end{abstract}

\maketitle

\renewenvironment{bnf*}%
{\csname align*\endcsname}%
{\csname endalign*\endcsname\ignorespacesafterend}

\renewenvironment{bnf}%
{\csname align\endcsname}%
{\csname endalign\endcsname\ignorespacesafterend}

\newcommand{\bnftsfont}[1]{\texttt{#1}}
\renewcommand{\bnfpo}{::=}
\newcommand{\bnfalt}{&\bnfor}
\renewcommand{\bnfts}[1]{\;\textnormal{\bnftsfont{#1}}\;}
\renewcommand{\bnfprod}[2]{\bnfpn{#1} \; & \bnfpo \; #2}

\definecolor{lightgray}{gray}{0.92}

\definecolor{javared}{rgb}{0.6,0,0} 
\definecolor{javagreen}{rgb}{0.25,0.5,0.35} 
\definecolor{javapurple}{rgb}{0.5,0,0.35} 
\definecolor{javadocblue}{rgb}{0.25,0.35,0.75} 

\lstset{language=C++,
basicstyle=\ttfamily,
keywordstyle=\color{javapurple}\bfseries,
stringstyle=\color{javared},
commentstyle=\color{javagreen},
morecomment=[s][\color{javadocblue}]{/**}{*/},
tabsize=4,
showspaces=false,
showstringspaces=false,
frame=single,
breaklines        = true,
breakatwhitespace = false,
breakindent       = 2ex,
escapeinside={(-}{-)},          
morekeywords={Tensor, size_t, to, IndexLabel,range},framexrightmargin=-3pt
}


\newcommand{\inlinecode}[1]{
\hspace{-.5em}
\colorbox{lightgray}{
\hspace{-.5em}\lstinline[basicstyle=\scriptsize\ttfamily]$#1$\hspace{-.5em}
}
\hspace{-.5em}
}
\newcommand{\code}[1]{{\ifmmode{\texttt{#1}}\else$\texttt{#1}$\fi}}
\newcommand{\boldit}[1]{{\ifmmode{\textit{\textbf{#1}}}\else$\textit{\textbf{#1}}$\fi}}

\section{Introduction}

The recent slowdown of growth in realized multi-core performance of commodity microprocessors has pushed vendors and users to consider more specialized architectures, including GPGPUs, FPGAs, and system-on-chip. 
Several domains, such as artificial intelligence, have experienced an explosion of highly-specialized heterogeneous architectures, including Google TPUs~\cite{jouppi2017datacenter}, NVIDIA DLA, and Intel Nirvana.
With such a large variety of architectures, performance portability and productivity have become as important as peak performance, if not more.
On one side, scientists and engineers have moved towards high-level, domain-specific (DSL) languages that facilitate implementing complex algorithms and allow them to focus on the algorithm's details rather than the specific idiosyncrasies of the underlying architectures.
On the other side, to achieve optimal performance on modern architectures, it is imperative to exploit hardware features by writing highly-specialized, low-level, architecture-dependent kernels. 

This struggle for balance is not a simple one to solve. A one-to-one mapping between each DSL and each architecture is impractical and expensive to maintain. Instead, researchers have looked into ways to abstract the domain-specific aspects of an implementation from the architecture-specific ones and identify intermediate representations (IRs) to realize such abstractions. 
For example, LLVM~\cite{llvm} maps multiple front-end programming languages (e.g., C, C++, and Fortran) to LLVM IR, and then maps this IR to various target architectures.
%
On the other hand, a generalized IR also means that domain-specific information and semantics are lost while lowering the code. 
For example, there is no simple way to express in LLVM IR common operations such as generic matrix-matrix multiplication (GEMM) or 2D convolution.
It follows that domain-specific optimizations are impractical to be performed at such low-level IR, which could introduce performance loss. 
To overcome this limitation, modern systems
(e.g., TensorFlow~\cite{tensorflow2015-whitepaper}, Rust, and Halide~\cite{halide2013}) propose high-level IRs where domain-specific optimizations are performed before lowering the code to the lower IRs. 

In this work, we introduce \name{}, a novel compiler infrastructure and programming language for tensor algebra targeting 
high-performance computational chemistry. \name{} increases productivity by providing very high-level programming abstractions that resemble Einstein notations~\cite{doi:10.1002/andp.19163540702}, performs sophisticated domain-specific code optimizations and rewriting, and generates code 
amenable to be executed on heterogeneous architectures. \name{} is based on the Multi-Level Intermediate Representation (MLIR) recently introduced by Google to simplify writing new compiler infrastructures. 
In the \name{} multi-level IR, domain-specific, application-dependent optimizations are performed at higher levels of the IR stack where operations resemble programming languages' abstractions 
and can be optimized based on the operations semantics. 
Generic, architecture-specific optimizations are, instead, performed at lower-levels, where simpler operations are mapped to the memory hierarchy and to processor's registers. 
A distinct advantage of a compiler-based approach compared to library-based solutions~\cite{BLIS_2015,mkl2012,hptt_17} is that \name{} can handle arbitrary multi-operand tensor expressions and perform state-of-the-art 
algorithmic and code optimizations. 
Our performance results indicate that the code automatically generated by \name{} is on par or better than manually-optimized 
tensor contractions from TCCG~\cite{SpringerTCCG2018} that leverage state-of-the-art computational libraries, such as BLIS~\cite{BLIS_2015} and HPTT~\cite{hptt_17}, and achieves up to $1.98\times$ speedup over a set of 30 contractions.
\name{} provides enough expressiveness and completeness to implement two complex methods (coupled-cluster \texttt{singles} and \texttt{doubles} excitation equations) from the NorthWest Chemistry (NWChem)~\cite{nwchem} computational chemistry package, which consists of 154 expressions and a total of 417 tensor contractions, and achieves up to $23.9\times$ speedup over executing tensor contraction in the natural order of the expression.
Additionally, by pairing our compiler to hardware accelerator simulators,
\name{} 
can be used to study novel data-parallel hardware designs  
 and their impact on the entire chemistry methods. 
To the best of our knowledge, \name{} is the first modular compiler framework that allows researchers to express complex tensor expressions, perform domain- and architecture-specific code optimizations and transformations outperforming hand-tuned solutions, and can be used 
for hardware-software co-design. 
In summary, we make the following contributions:
\begin{itemize}
\item \name{}, a novel compiler and DSL for tensor algebra that specifically targets chemistry applications with support for multi-operand expressions;
\item a multi-level IR, a set of progressive lowering steps, and a series of domain- and architecture-specific optimizations to generate efficient code; 
\item a new co-design methodology to study custom accelerators;
\item a comparison with state-of-the-art, manually-implemented tensor contraction benchmarks that leverage highly-optimized computational libraries.
\end{itemize}


\section{Tensor Contractions}
\label{sec:background}
Tensor contractions are high-dimension analogs of matrix multiplications widely used in many scientific and engineering domains, including deep learning, quantum chemistry, and finite-element methods. For example, the perturbative triples correction in couple cluster CCSD(T)~\cite{ccsd_t} methods used in the NWChem computational chemistry framework~\cite{nwchem} originates a 6D output tensor from two 4D inputs tensors. Tensor contractions are computationally intensive
and dominate the execution time of many computational applications.

Because of their wide applications in many fields, tensor contractions have been widely studied and optimized. Consider the following tensor contraction, expressed using Einstein notation~\cite{doi:10.1002/andp.19163540702}, where two 4D tensors, $A$ and $B$, are contracted to produce a 4D tensor $C$:
\begin{equation}
C[a,b,c,d] = A[a,e,b,f] * B[d,f,c,e]
\label{eq:contraction}
\end{equation}

\noindent In this contraction, the indices $e, f$ appear in both right-hand tensors but not in the left-hand tensor $C$ (\emph{summation} or \emph{contraction} indices). 
The indices $a, b, c, d$ appear in exactly one of the two input tensors and the output tensor (\emph{external} or \emph{free} indices). 
A tensor contraction is, thus, the contraction of the two input tensors $A$ and $B$ over the contraction indices $e, f$:\\
\begin{equation}
C[a,b,c,d] = \sum_{e,f}{A[a,e,b,f] * B[d,f,c,e]}
\end{equation}

\noindent A \naive way to perform the above computation is to directly lower to a nested-loop implementation of the problem. Such implementations have been shown to be inefficient due to poor data locality. A more efficient approach, commonly used in modern high-performance tensor libraries, leverages highly optimized GEMM engines.
This approach, often referred as transpose-transpose-GEMM-transpose (TTGT), performs the permutations of the input tensors followed by a high-performance matrix-matrix multiplication and a final permutation to reconstruct the output tensor.
The first two transposes ``flatten'' a multi-dimensional tensor into a 2D matrix by first permutating the indices so that they are contiguous in memory ($A \rightarrow TA$) and then merging pairs of consecutive indices to form lower-dimensional tensors ($(a,b) \rightarrow i$, $(e,f) \rightarrow j$, $(d,c) \rightarrow k$):

\begin{align*}
A[a,e,b,f] &\rightarrow TA[a,b,e,f] = A_p[i,j];\\
B[d,f,c,e] &\rightarrow TB[e,f,d,c] = B_p[j,k] \notag
\end{align*}

\noindent The tensor contraction expressed in equation \ref{eq:contraction} can then be expressed as
\begin{equation}
C_p[i,k] = A_p[i,j] * B_p[j,k] 
\end{equation}
%
%
where $C_p[i,k] = TC[a,b,d,c] \rightarrow C[a,b,c,d]$. 
The TTGT method is effective to perform high-efficient tensor contractions despite the overhead of potentially performing three additional permutations. In fact, highly-optimized GEMM operations perform considerably better than nested loop implementations on modern architectures and exploit high data locality (see Section~\ref{sec:opts}).
In this work, we consider employing custom accelerators to perform even more efficient GEMM operations, thus our compiler produces target code that is optimized and amenable to such accelerators (Sections~\ref{sec:opts} and~\ref{sec:mod}).

\section{The \name{} Compiler Infrastructure}
\label{sec:compiler}

\begin{figure}[t]
  \begin{center}
    \includegraphics[width=0.50\textwidth]{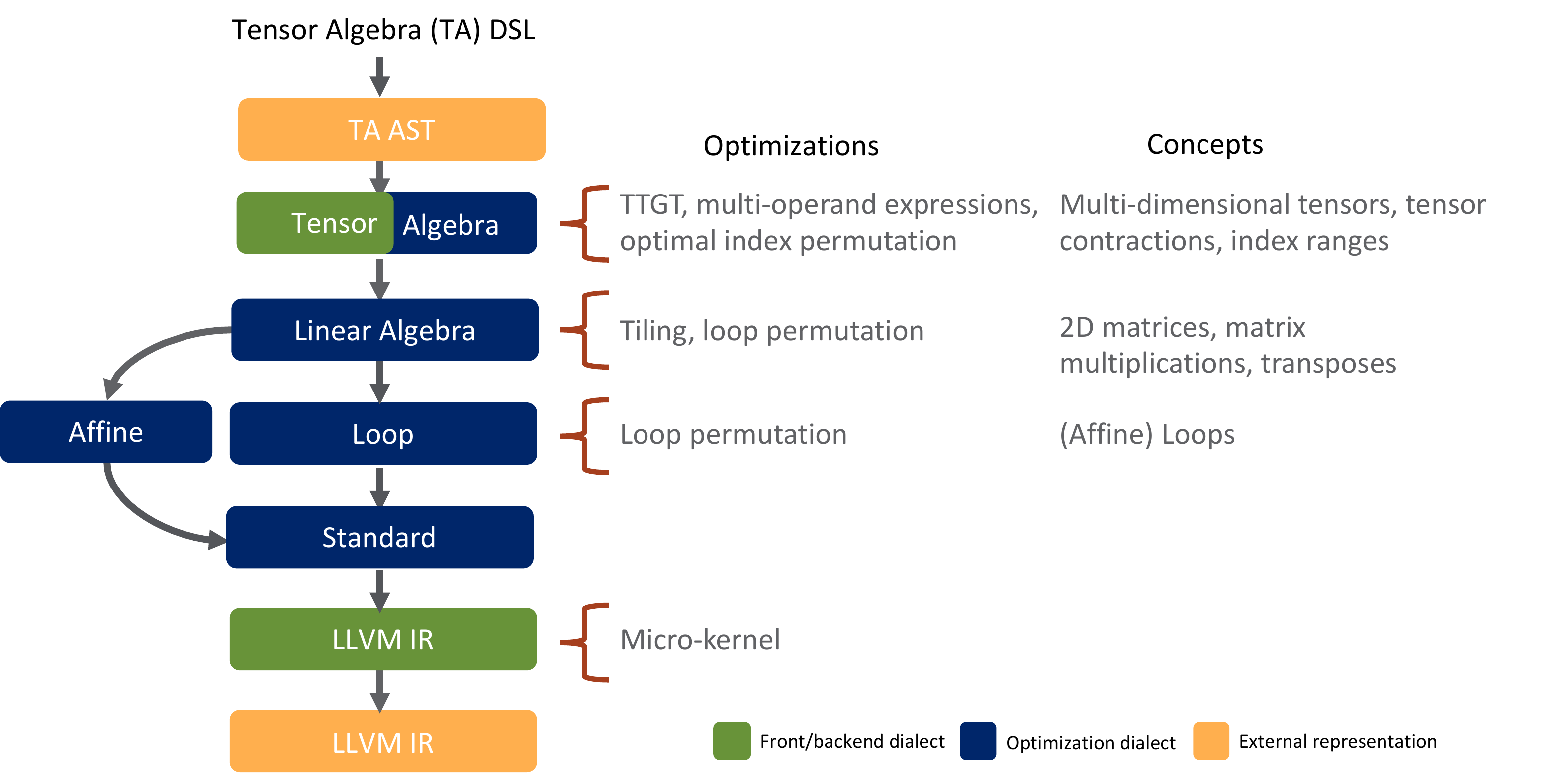}
  \end{center}
  \caption{\name{} execution flow and compilation pipeline}
  \label{fig:overview}
\end{figure}

\begin{figure*}[t]
\begin{subfigure}[b]{0.4\textwidth}
        \begin{bnf*}
          \bnfprod{ilabel}
            {\bnfts{IndexLabel} \\ &\phantom{::=} \bnfpn{id-list} \bnfts{=} \bnfpn{range} \bnfts{;}}\\
          \bnfprod{range}
            { \bnfts{[} \bnftd{int} \bnfts{]} \\ \bnfalt 
              \bnfts{[} \bnftd{int} \bnfts{:} \bnftd{int} \bnfts{:} \bnftd{int} \bnfts{]}}\\
          \bnfprod{tensor}
            { \bnfts{Tensor} \\ &\phantom{::=} \bnfts{<} \bnfpn{element-type} \bnfts{>} \bnfpn{id-list} \bnfts{;}}\\
          \bnfprod{id-list}
            {\bnfpn{id} \bnfor \bnfts{[} \bnfpn{id} \bnfts{,} \bnfpn{id-list} \bnfts{]}}\\
          \bnfprod{id-list}
            {\bnftd{int} \bnfor \bnftd{double} \bnfor \bnftd{float}}\\  
          \bnfprod{id}
            {\bnftd{any object identifier}}
        \end{bnf*}
    \caption{Tensor label grammar}
    \label{fig:tensor_const}
\end{subfigure}
\begin{subfigure}[b]{0.4\textwidth}
    \begin{bnf*}
        \bnfprod{tensor-op}
            {
              \bnfpn{op-lhs} \bnfts{=} \bnfpn{op-rhs} \\ 
              \bnfalt \bnfpn{op-lhs} \bnfts{+=} \bnfpn{op-rhs}\\
              \bnfalt \bnfpn{op-lhs} \bnfts{-=} \bnfpn{op-rhs}
            }\\
          \bnfprod{op-lhs}
            {\bnfpn{label-tensor}}\\
          \bnfprod{op-rhs}
            {
              \bnfpn{alpha} \bnfor \bnfpn{label-tensor} \\ \bnfalt 
              \bnfpn{alpha} \bnfts{*} \bnfpn{label-tensor} \\ \bnfalt
              \bnfpn{alpha} \bnfts{*} \bnfpn{label-tensor} \bnfts{*} \bnfpn{label-tensor} \\ \bnfalt
              \bnfpn{op-rhs} \bnfts{*} \bnfpn{label-tensor}
            }\\
          \bnfprod{label-tensor}
            {\bnfpn{tensor-id} \bnfts{(} \bnfpn{label-list} \bnfts{)}}\\
          \bnfprod{alpha}
            {\bnftd{tensor value type}}
        \end{bnf*}
    \caption{Tensor operations grammar}
    \label{fig:tensor_ops}
\end{subfigure}
\caption{Tensor label and operation grammar.}
\label{fig:ta_grammer}
\end{figure*}

Our proposed compiler infrastructure consists of a DSL language for tensor algebra computations, a progressive lowering process to map high-level operations 
to low-level architectural resources, 
a series of optimizations performed at each step in the lowering process, and various IR dialects to represent key concepts, operations, and types at each level of the multi-level IR.
\name{} is based on the MLIR framework~\cite{lattner2020mlir}, a compiler infrastructure to build reusable and extensible compilers. MLIR supports the compilation high-level abstraction and domain-specific constructs and provides a disciplined, extensible compiler pipeline with gradual and partial lowering. 
Users can build domain-specific compilers and customized IRs, as well as combining existing IRs, opting in to optimizations and analysis. 


Figure~\ref{fig:overview} shows the compilation pipeline of \name{}. Users express their computation using a high-level tensor algebra language 
(Section~\ref{sec:language}). The language operators, types, and structures are first mapped to an abstract syntax tree and then to the tensor algebra (TA) \emph{dialect}, the first dialect in the \name{} IR stack. The TA dialect contains domain-specific concepts, such as multi-dimensional tensors, contractions, and tensor expressions. Here, several domain-specific optimizations are performed, such as reformulating tensor contractions using the TTGT method. 
Next, \name{} lowers the TA dialect representation of the computation to a mixed dialect based on the linear algebra (LinAlg) and Affine loop dialects. 
High-level concepts, such as tensor contractions, are replaced with more general operations (transpose, matrix multiplication, affine maps, etc.).
At this stage, there is a departure from domain-specific concepts but the operations are still architecture-independent. The next step consists of further lowering of LinAlg operations 
to the (Affine) Loop dialect: at this stage, \name{} performs architecture-specific optimizations and requires information for the specific target. GEMMs are tiled to fit matrix slices in the processor's data caches 
as well as to map computation to the processor's registers. The innermost GEMM computation is performed using an architecture-specific, highly-optimized micro-kernel (Section~\ref{sec:opts}) or (simulated) custom accelerators (Section~\ref{sec:mod}).
Finally, \name{} lowers the program to standard dialect and then to the LLVM dialect, 
which is then mapped to LLVM IR and lowered to machine instructions for execution.

\section{Tensor Algebra Language}
\label{sec:language}

\begin{figure*}[t]
    \begin{subfigure}{0.3\textwidth}
        \includegraphics[width=\textwidth]{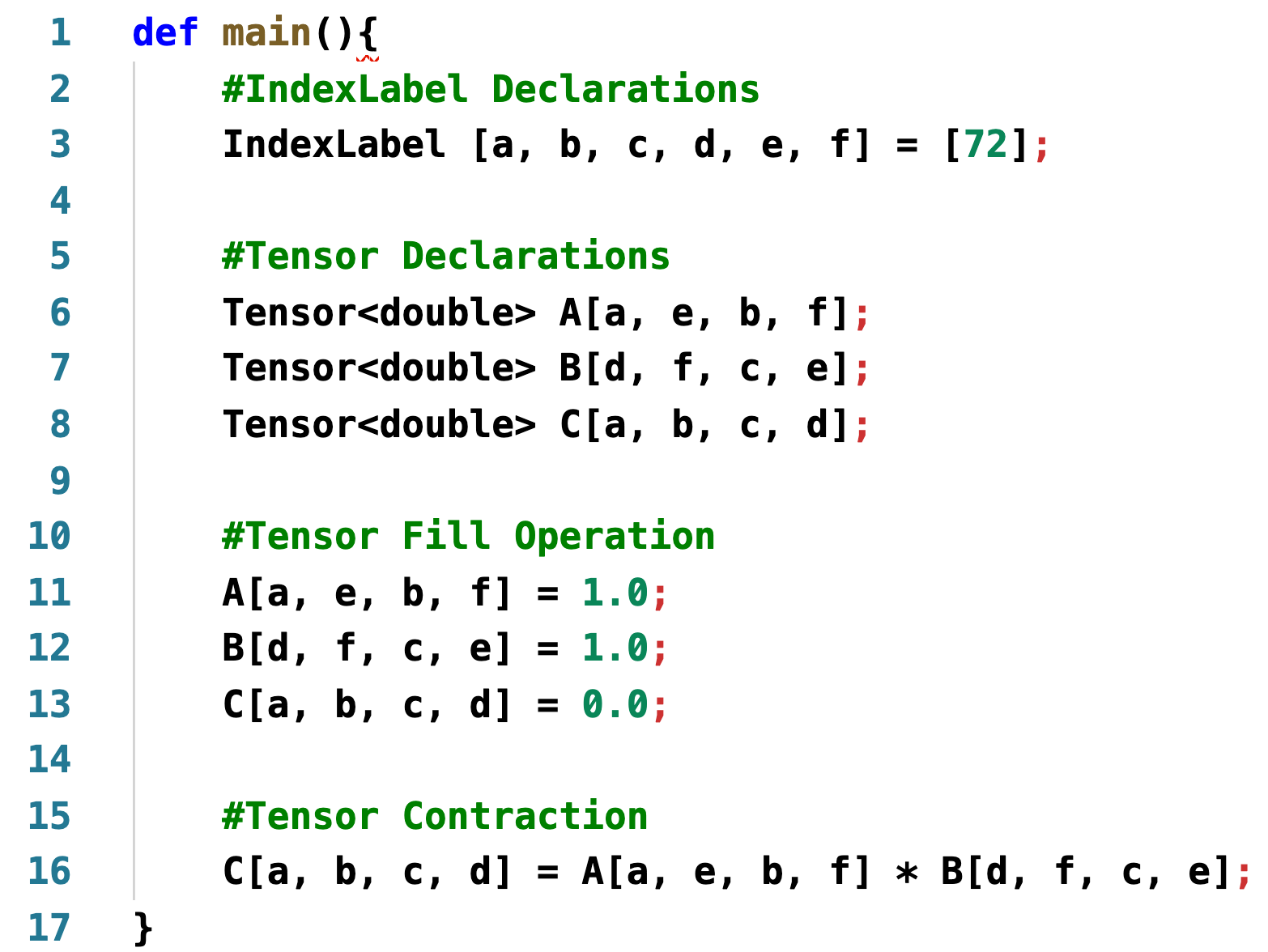}
        \vspace{10em}
        \caption{DSL TA language}
        \label{fig:ta_language}
        \vspace{-2em}
    \end{subfigure}
    \begin{subfigure}{0.60\textwidth}
        \includegraphics[width=0.9\textwidth]{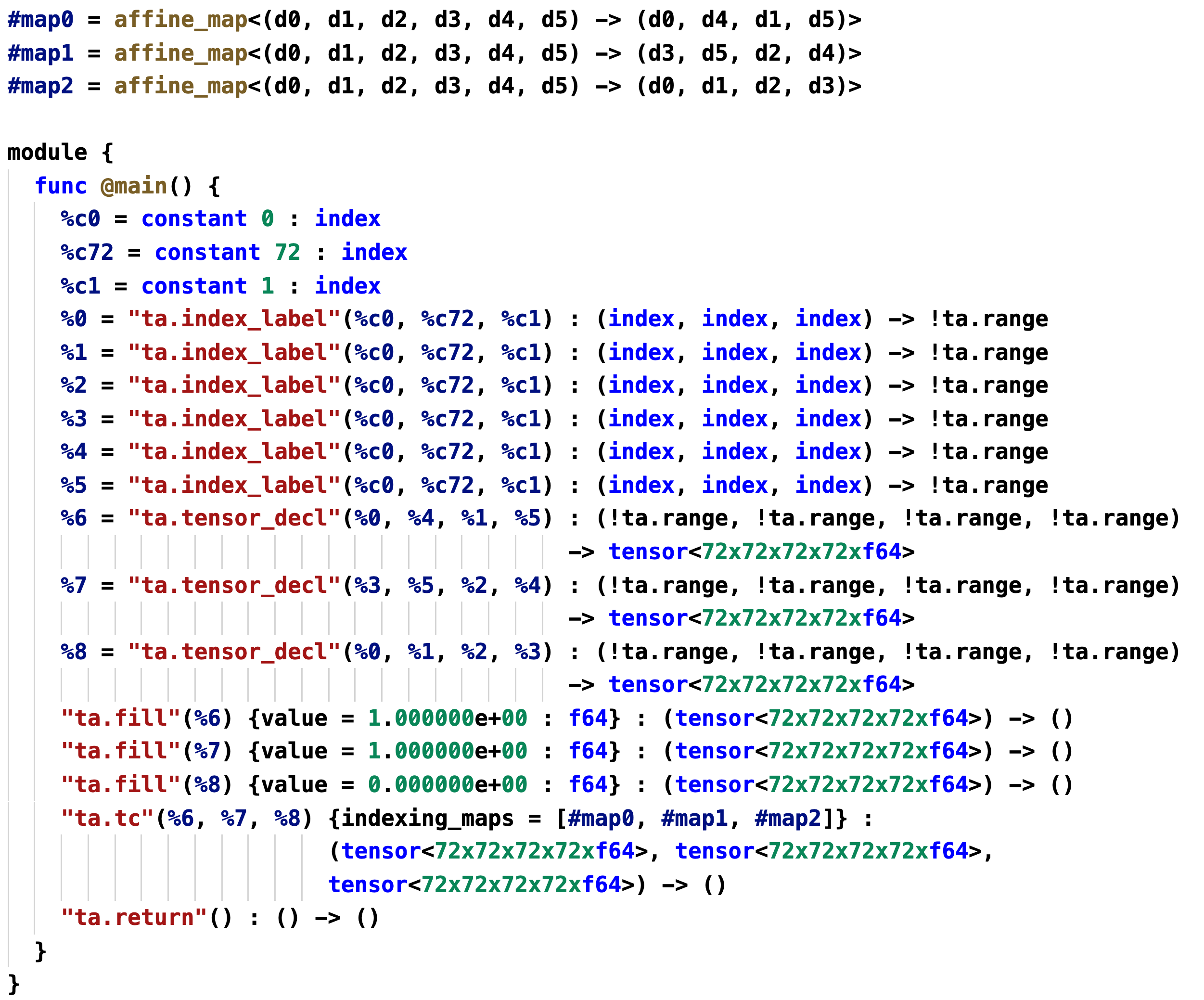}
        \caption{TA dialect}
        \label{fig:ta_dialect}
    \end{subfigure}
    \caption{Example tensor contraction in \name{} DSL and its relative TA dialect.}
\end{figure*}

We developed a high-level Tensor Algebra (TA) DSL for tensor algebra computation. As for any DSL, the goal of the \name{} language is to allow scientists 1) to express concepts and operations in a form that closely resembles familiar notations and 2) to convey domain-specific information to the compiler for better program optimization. 
Our language represents Einstein mathematical notation and provides users with an interface to express tensor algebra semantics. 

Figure~\ref{fig:tensor_const} describes the tensor structures and how they are represented and constructed in our DSL. A tensor object refers to a multi-dimensional array of arithmetic values that can be accessed by using indexing values. 
Range-based index label constructs ($\bnfpn{ilabel}$) represent the range of indices expressed through a scalar, a range, or a range with increment. Index labels can be used both for constructing a tensor ($\bnfpn{tensor}$) or for representing a tensor operation ($\bnfpn{label-tensor}$). In a tensor construction, index labels are used to represent each dimension size. 
In the context of a tensor operation, they represent slicing information of the tensor object where the operation will be applied.
Figure~\ref{fig:tensor_ops} shows the grammar for supported tensor operations. Similar to the tensor construction, index labels are used as the main construct for representing the operations on the tensors. 
Each tensor operation production rule ($\bnfpn{tensor-op}$) is composed of a left-hand side (lhs) and a right-hand side (rhs) operation. While lhs can only be a labeled tensor construct, rhs can be of different types:
\begin{itemize}
  \item alpha value (\texttt{A[i,j] = <const>}), which corresponds to a tensor fill operation where all the elements in the tensor are set to a scalar value.
  \item labeled tensors (\texttt{A[i,j] =  alpha * D[j,i]}) correspond to a tensor copy operation (with respect to the label permutation). If the index label used in the rhs tensor is different from the one using during the tensor construction, the lhs tensor will represent a \emph{view}.
  Following tensor operations that employ this lhs tensor will operate on a slice of the original tensor.
  \item multiplication of two labeled tensors (\texttt{C[i,k,l] =  alpha * A[i,j] * B[j,k,l]}) 
  updates the lhs with the tensor contraction results. 
  \item multi-operand expressions (\texttt{D[i,l] = alpha * A[i,j] * B[j,k] * C[k,l]}) computes the whole pipeline of tensor contractions and updates the lhs.
\end{itemize}


The \name{} TA language simplifies writing tensor algebra program by supporting common programming paradigms and enables users to express high-level concepts with familiar notations. Figure~\ref{fig:ta_language} shows a general tensor contraction implementation using \name{} TA language. Line 3 describes an index label representing the size of each tensor dimension and the operation labels for describing the tensor fill and tensor contraction. Our TA language supports defining multiple \texttt{IndexLabel} variables (similar to structured bindings in C++-17) in a single statement.  Tensors are then constructed using the \texttt{IndexLabel}s and the element type (lines 6-8). 
and contracted over indices \texttt{[e,f]} (line 16). 
\section{The Tensor Algebra Dialect}
\label{sec:dialect}


The Tensor Algebra dialect is the first dialect in the~\name{} compiler stack (Figure~\ref{fig:overview}). The main goal of this dialect is to represent basic building blocks of the tensor algebra computation, describe tensor algebra specific types and operations, and represent semantics information expressed through the TA DSL. 
Figure~\ref{fig:ta_dialect} shows the TA dialect representation of the tensor contraction example shown in Figure~\ref{fig:ta_language}.
We define new operations in TA dialect that correspond to each tensor algebra DSL operation semantic. A \texttt{ta.index\_label} corresponds to an \texttt{IndexLabel} construct in the TA language while a \texttt{ta.labeled\_tensor} is for the \texttt{LabeledTensor} constructs. 
Figure~\ref{fig:ta_dialect} shows how an index label operation is constructed using 
a range (\texttt{!ta.range}) type. 
New tensors are declared with the \texttt{ta.tensor\_decl} operation, which takes as input the index labels for each dimension and the data type.
The \texttt{ta.labeled\_tensor} operation represents a slice of a tensor that is being used in any operation that references it. 
This operation takes a tensor declaration and a set of index label references as inputs to construct a sliced version of the tensor.


Three classes of tensor operations are currently supported: unary (fill, copy, set), binary (contraction), and multi-operand expressions (contraction chains). \texttt{ta.fill} initializes a tensor object with a single value provided as an attribute to the operation. \texttt{ta.copy} performs an element-wise copy operation between two tensors scaling the output tensor by factor \texttt{alpha}.
\texttt{ta.set} operates similarly to \texttt{ta.copy} but takes as input the result of a binary operation instead of a tensor. This operation is used to support multi-operand tensor contractions. 
Tensor contractions \texttt{ta.tc} 
take as input the input and output tensors, the scaling value \texttt{alpha}, and the \texttt{indexing map}s for the labels used in the contraction. 

For multi-operand expressions that involve several contractions, we introduce a utility operation (\texttt{ta.mult}) that represents a binary operation. The actual computation for a multi-operand expression includes calculation of intermediates and then the actual tensor contractions. We represent the order of binary operations with a binary tree and assign results to the output tensor using \texttt{ta.set}. 
\br{Any simple illustration of this?}

\section{Optimizations and Transformations}
\label{sec:opts}


The main advantage of a multi-level IR is that different kinds of optimizations can be applied at each level of the IR stack and optimizations can be shared and reused across different stacks. In the \name{} compiler, we apply domain-specific optimizations at the TA dialect level, general optimizations at the LinAlg dialect level, and architecture-specific optimizations at the lower levels. In the following, we explain our optimizations from the top IR level while lowering the code. 

\begin{figure*}[t]
    \centering
    \includegraphics[width=0.8\textwidth]{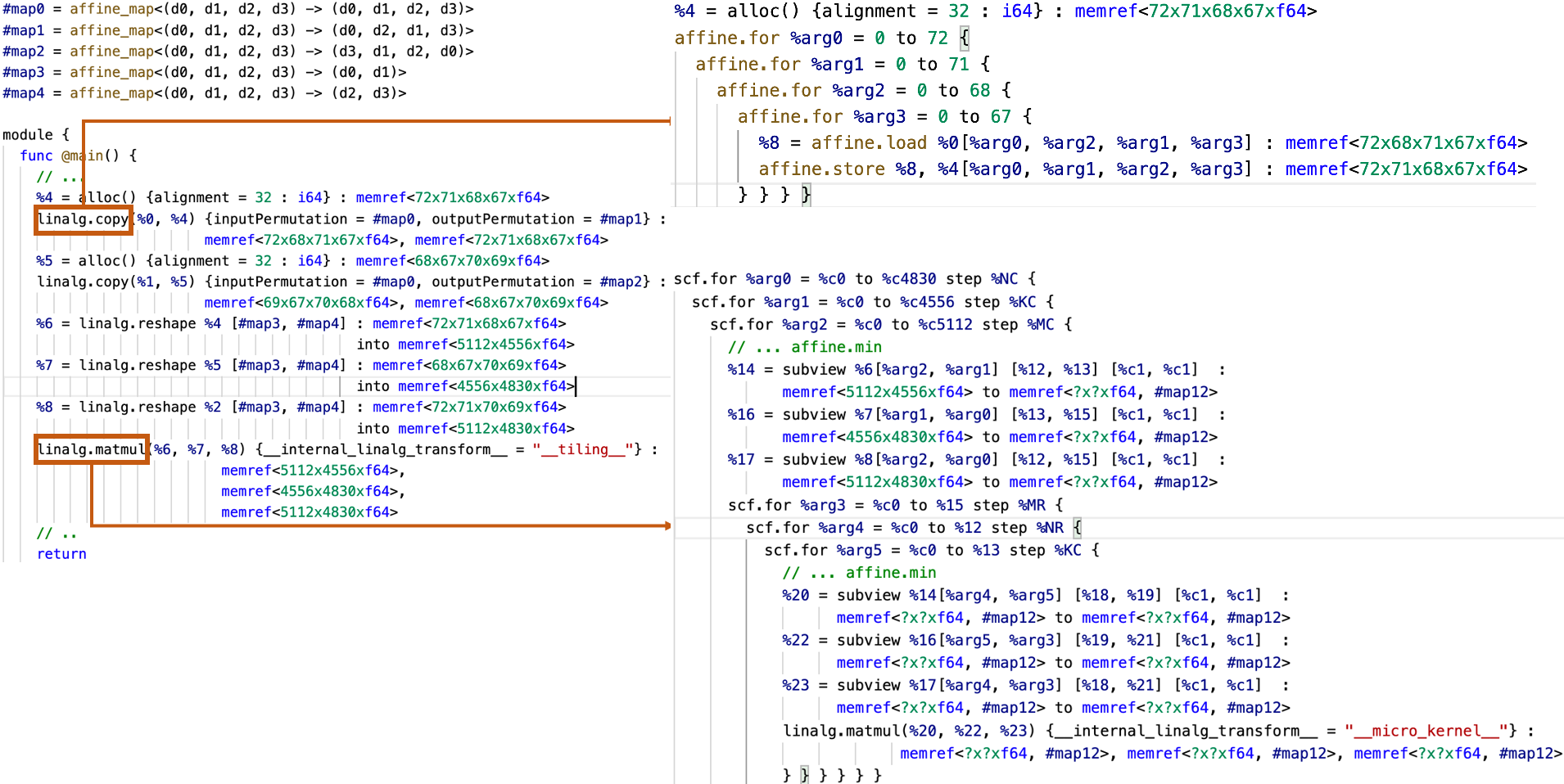}
    \caption{
    TA dialect after reformulation with a TTGT method (left panel). Lowering and optimization of transpose (right-top) and GEMM (right-bottom).}
    \label{fig:ta_lowering_opt}
\end{figure*}

\textbf{TTGT} As discussed in Section~\ref{sec:background}, tensor contractions can be reformulated by transposing multi-dimensional input tensors into 2D matrices, performing a GEMM operation, and unflatting the output tensor back to its original form. Although, this approach incurs the additional overhead of transpose operations, employing highly-optimized GEMM kernels outweighs this overhead. The left panel in Figure~\ref{fig:ta_lowering_opt} shows the TTGT reformulation of the \texttt{ta.tc} operation: transpose of input tensors (\texttt{linalg.copy}) and a GEMM operation (\texttt{linalg.matmul}).

\textbf{Optimal permutation for TTGT} 
The  permutation chosen to reformulate a tensor contraction using the TTGT method has a considerable impact on performance.
The cost of transposing a high-dimension tensor into a 2D tensor depends on which indices are transposed and the storage format. For row-major format, transposing the first indices is more expensive than transposing the latest ones, especially for the output tensor.
In order to select the best permutation, we use a cost model based on a heuristic that assigns higher costs to permutations that move the outermost dimensions. Additionally, some permutation naturally results in a reduction of the number of transpose operations. 
We compute the cost of each valid transposition of input and output tensors, including the position swap for the input tensors, and select the permutation with the lowest cost.

\textbf{Multi-Operand Expression} Given the associative property of tensor contractions, the order in which  contractions are computed in a multi-operand expression produces the same results. However, the \emph{number of operations} performed may vary depending on the grouping order of the contractions. Performance variation may be significant, especially if some of the tensors involved have low cardinality in some dimensions (e.g., ``skinny matrices''). We explore all possible ordering of a multi-operand expression and identify the one that minimizes the overall number of operations. 
We then organize the sequence of operations in a binary tree and lower the multi-operand expression to a sequence of tensor contractions. Note that because the shape of the intermediate tensors is different from the original one, some tensor contractions may degenerate to simpler lower-dimension operations, such as GEMM or tensor-vector multiplications, which are further optimized (e.g., removing additional transpose).

\textbf{Transpose optimization} 
Our transpose optimization consists of two steps: loop permutation and tiling.
The main rational of the cost model is that the loops corresponding to the innermost indices should be at the innermost level. 
We assign a weight to each loop index according to its position in the input and output tensor (the weight is higher if an inner index does not correspond to an inner loop) and compute the overall cost of the permutation by summing these weights. 
The right-top panel in Figure~\ref{fig:ta_lowering_opt} shows the IR after the selected loop permutation ($i,j,k,l \rightarrow i,k,j,l$). 
Next, we employ tiling to improve  locality. 

\textbf{GEMM optimizations}
GEMM plays a paramount role in a broad range of domains such as deep learning, scientific computing, and image processing. Tiling and blocking are effective methods to improve data locality and overlap computation and memory access. We employ the same tiling strategy used~\cite{BLIS_2015, mkl2012}: given $C[M,N] = A[M,K] * B[K,N]$, 
the $C$ matrix is partitioned into multiple tiles with size $M_c\times N_c$. Each tile of $C$ needs to access to $A$ matrix with size $M_c\times K$ and a whole column of $B$ matrix with size of $K\times N_c$. Since the whole row band of $A$ and column of $B$ are still large to be accommodated in the processor's cache, the $K$-dimension has to be partitioned into smaller tiles of $A$ ($M_c\times K_c$) and $B$ ($K_c\times N_c$).
$M_c$, $K_c$ and $N_c$ are carefully selected so that the sub-matrix of $B$ ($K_c\times N_c$) fits into the L3 cache and the $A$ sub-matrix ($M_c\times K_c$) fits into the L2 cache. The $N_c$ and $M_c$ dimensions are further tiled by  $N_r$ and $M_r$, respectively, so that the sub-matrix of $A$ ($M_r\times K_c$) and $B$ ($K_c\times N_r$) fit into the L1 cache.
The $M_r\times N_r$ elements from the matrix $C$ fit into registers and the innermost computation of size $M_r\times N_r$ is executed as a micro-kernel. The right-bottom panel in Figure~\ref{fig:ta_lowering_opt} shows the various levels of tiling. 

\textbf{GEMM micro-kernel} Modern architectures are very complex and require sophisticated and highly-optimized code to fully achieve high performance. Compiler frameworks do their best to generate such code but also remain general. 
\name{} leverages the LLVM back-end for code and binary generation which, when combined with tiling optimizations, provide good performance. However, a highly-optimized code for the specific architecture can fully leverage vector instructions, instruction-level parallelism, speculation, and other architectural features, achieving even higher performance. Also, \name{} has been designed to support custom hardware accelerators that implement specific functionalities in hardware. Thus, the innermost computation in the GEMM kernel is performed using a \emph{micro-kernel}, either a highly-specialized code for the target architecture (``soft-accelerator'') or a custom hardware accelerator (``hard accelerator''). 

\section{Modeling Custom Accelerators}
\label{sec:mod}
Custom hardware accelerators may significantly increase performance, area, and energy efficiency of high-end compute systems. It comes as no surprise that hardware acceleration is widely employed in many domains, including mobile, automotive, high-performance computing, and machine learning. However, designing custom accelerators requires computationally-intensive simulations using software or FPGA-based tools, 
which may limit the scope to very small kernels.

\name{} provides an opportunity to perform  co-design and design space exploration (DSE) efficiently and to assess the performance of the entire application, instead of only the innermost kernel. As we explained in the previous section, the lower level dialects 
represent architecture-specific operations and the innermost computation in the GEMM operation is implemented as an architecture-specific, highly-optimized micro-kernel. In order to perform co-design for a target accelerator and measure the impact of realistic tensor contractions, we replace the micro-kernel with a timing model of the hardware accelerator and execute the entire contraction at native speed. To this extend, we pair \name{} with Aladdin~\cite{aladdin2014}, a pre-register-transfer level (RTL), power-performance simulation framework that targets rapid prototyping of data parallel accelerators. Aladdin specifications are essentially C representations of the functionalities that need to be implemented in hardware. From these representations, an LLVM-based tracer extracts a dynamic data dependence graph (DDDG) that describes the accelerator. 
Next, Aladdin applies various optimizations and resource constraints, 
therefore generating a realistic design.
Finally, Aladdin estimates power and performance from dynamic traces obtained from a driver program. 

Performing DSE for the hardware accelerator design with Aladdin takes order of minutes (instead of hours as in traditional FPGA-based DSE) while executing tensor contraction with \name{} is performed at native speed. The entire process, thus, can be completely automated and executed within minutes. 


\section{Evaluation}
\label{sec:evaluation}
This section presents the performance code generated by the \name{} compiler from the TA language. 
First, we show the performance impact of progressively applying the various optimizations described in Section~\ref{sec:opts}. Next, we compare our automatically-generated code against the code that leverages hand-optimized libraries. Finally, we show a co-design case for the GEMM accelerator. 
We perform our experiments on a compute node featuring Intel Xeon 6126 CPU at 2.60GHz and 192 GB of memory. We compare our results to 30 tensor contractions from the TCCG benchmark suite~\cite{SpringerTCCG2018}, using the reference problem sizes.  The first eight contractions include tensor-matrix multiplication from machine learning domain (1st to 8th). The next three contractions are used to transform a set of two-electron integrals from an atomic orbital basis to a molecular orbital basis (9th to 11th). Finally, the following 19 contractions are from the CCSD method of quantum chemistry. TCCG benchmarks are implemented in C++ and leverage highly-optimized computational libraries for tensor transpositions (HPTT) and GEMM kernels (BLIS). The results reported are the average of ten runs.

\begin{figure*}[t]
  \begin{center}
    \includegraphics[width=0.9\textwidth]{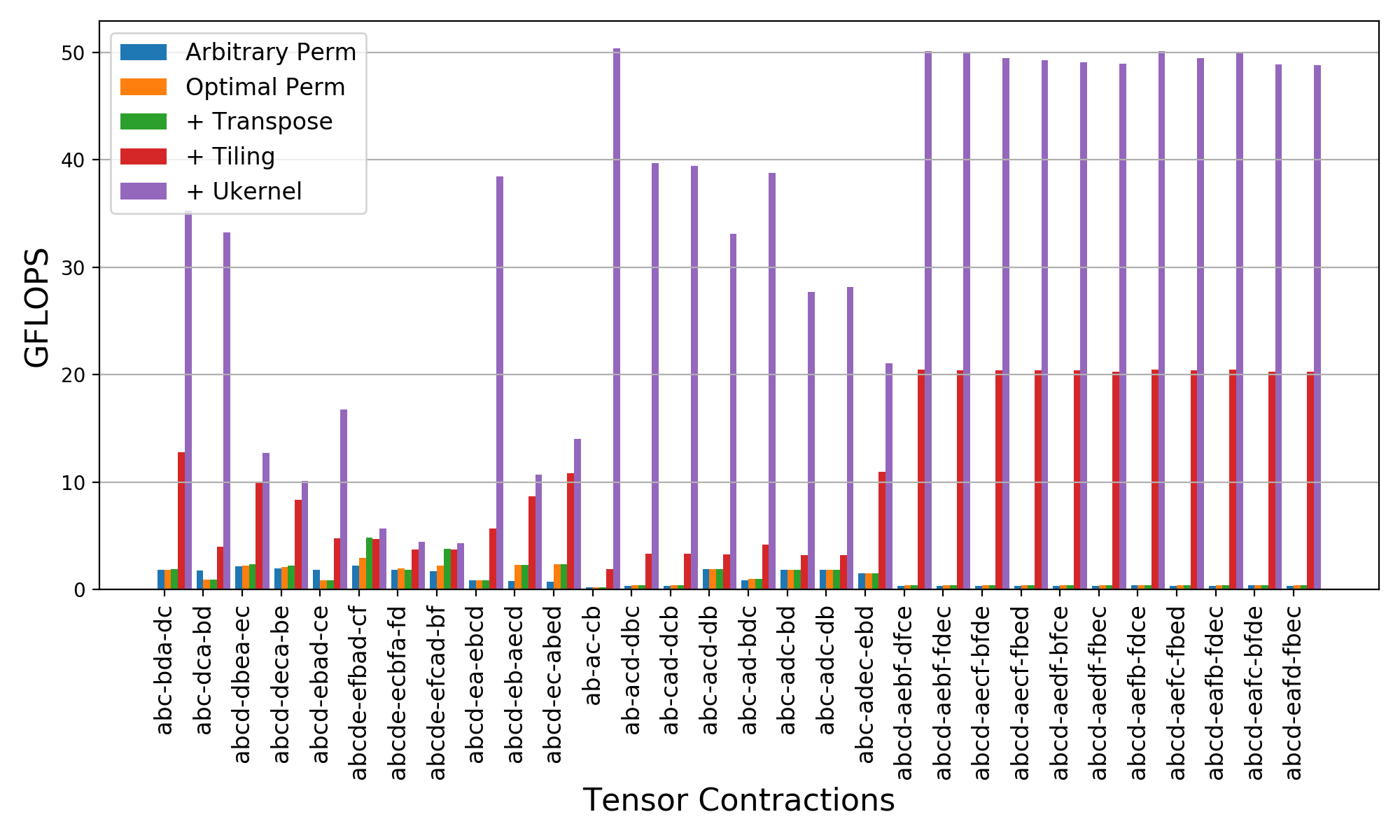}
  \end{center}
  \caption{Performance breakdown. The plot shows the impact of code optimizations applied incrementally to multiple tensor contractions (higher is better).}
  \label{fig:breakdown}
\end{figure*}

\begin{figure*}[t]
  \begin{center}
    \includegraphics[width=0.8\textwidth]{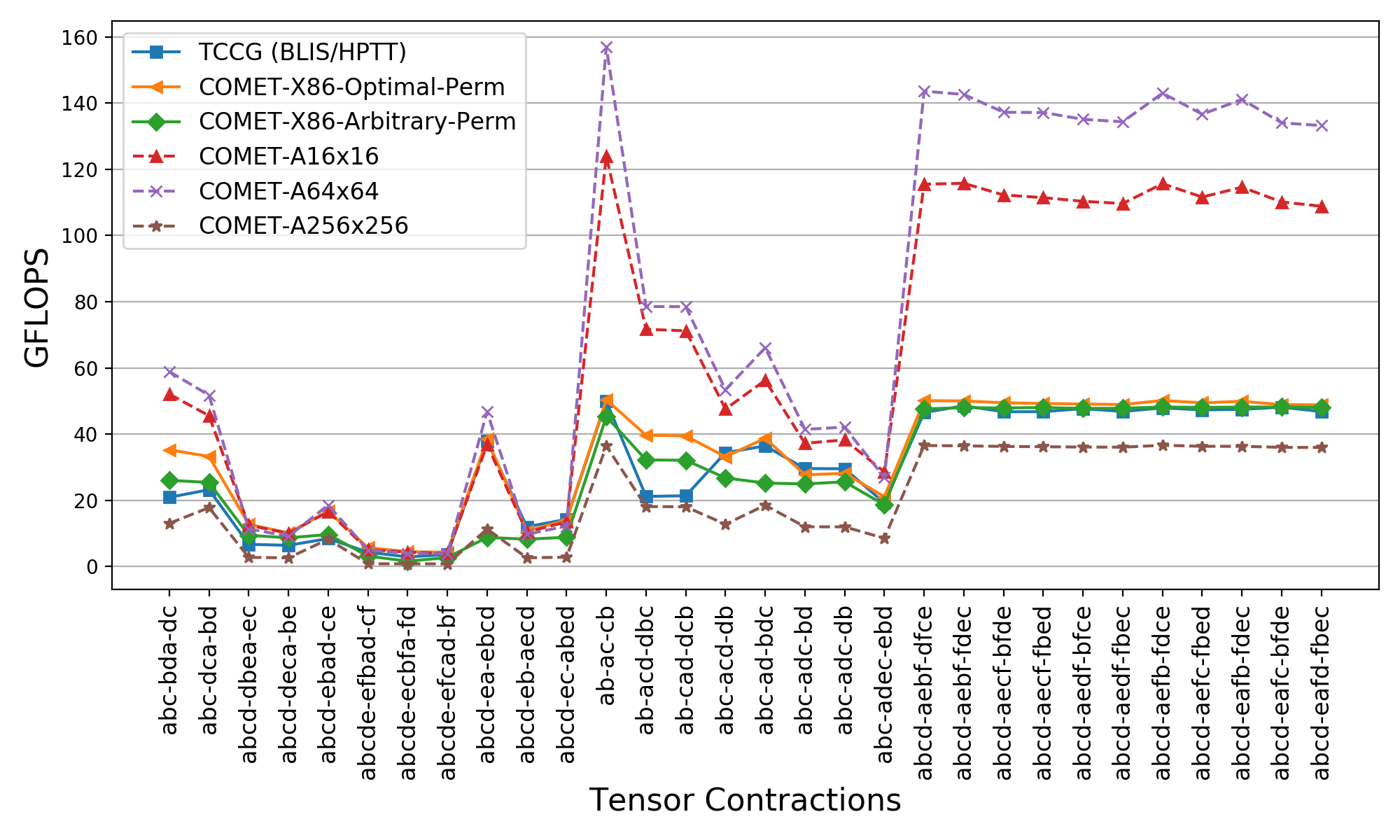}
  \end{center}
  \caption{\name{}'s performance comparison against the hand-optimized TCCG benchmarks on x86 (solid lines) and emulated platforms (dashed lines).
  }
  \vspace{-1em}
  \label{fig:comparison}
\end{figure*}

\textbf{Overall Performance Evaluation} Figure~\ref{fig:breakdown} shows the performance impact of applying the optimizations described in Section~\ref{sec:opts} to the TCCG tensor contractions written with \name{} DSL.
The x-axis represents each tensor contraction, and the y-axis denotes the performance in terms of GFLOPS.
We start from TTGT reformulations with a nested-loop GEMMs and transposes (blue bars) and progressively apply  architecture-independent 
and architecture-specific optimizations, 
thus each bar in the graph shows the incremental benefits. 
The plot shows that each optimization brings varied performance gains on the different tensor contractions. The architecture-independent optimizations may (or may not) bring immediate benefit.
This is the case of the arbitrary permutation for TTGT 
being already the optimal one.
However, in some cases, optimizing transpose operations almost double performance (6th and 8th contractions). It is important to note that some architecture-independent optimization, such as selecting the optimal permutation, might bring performance gains after all other optimizations are applied (Figure~\ref{fig:comparison}).
Architecture-specific optimizations, such as GEMM tiling, bring considerable performance improvements in most of the cases, achieving $23\times$ speedup on average and up to $56\times$ speedup compared to the equivalent loop nest. For high-dimensional tensor contractions (e.g., the ones from CCSD), for which data movement dominates the execution time, exploiting data locality provides significant performance improvements.
Finally, employing an architecture-specific micro-kernels that leverages AVX512 vector instructions, memory pre-fetching, and speculation, greatly improves performance, especially for the compute-bound contractions, achieving up to 51 GFLOPS. We remark that the micro-kernel is only effective once all other optimizations are performed, otherwise vector instructions cannot leverage data locality.


Figure~\ref{fig:comparison} compares \name{} against state-of-the-art implementation of TCCG tensor contractions. While both \name{} and the C++ implementations use the TTGT method, TCCG implementations (solid blue line) leverage highly-optimized computational libraries. \name{} implementations, instead, employ the optimizations described in the previous sections.
In order to perform a fair comparison, we use the same micro-kernel implemented by the BLIS library, hence both hand-tuned and \name-generated code use the same architecture-specific code.
The plot shows that \name{} results (solid yellow line) are comparable and, in some cases, better than the TCCG C++ implementations: \name{} achieves an average of $1.22\times$ and up to $1.98\times$ speedup compared to the TCCG benchmarks. 
However, we believe that \name{} has the distinct advantages of being more general (optimizations can be selectively applied), portable (different targets architectures may be chosen without re-implementing high-level optimizations), user-friendly (the TA language allows expressing  equations in mathematical forms), and does not force programmers to bend algorithms  to specific libraries APIs and formats.
The graph also shows \name{} performance when employing an arbitrary permutation instead of the best one (solid green line). Using the best permutation performs up to 4.36x better (1.38x on average) than using an arbitrary permutation. These results show that, although it is not evident at first (Figure~\ref{fig:breakdown}),  choosing an optimal permutation does increase the overall performance once architecture-specific optimizations are applied.

\gk{Jacques, does the current text address your question?}
\jp{I'm having some trouble interpreting the above, it sounds as if it says we could improve over the previous by replacing the inner kernels with optimized GEMM implementations. But then the graph seems to show a comparison with hand optimized code - so is the there a step in between these and previous results missing? E.g., DuoMD [which is named such in graph vs COMET] without optimized GEMM but corresponding to final bar of Fig 6? \gk{Correct, we can get better performance than in fig 5 by using a micro-kernel. The problem is that performance are so high that you can't read the graph anymore for the initial optimizations. So we thought about breaking the results in two: all optimizations except micro-kernel, and all optimizations including micro-kernel and comparing that with BLIS and the hardware accelerators, since they all use the same optimizations. Comet without micro-kernel, but with tiling, is the final bar in fig 5, fig 6 is the same but with micro-kernel}}

Overall, the results in Figures~\ref{fig:breakdown} and~\ref{fig:comparison} show that to achieve high performance it is imperative to employ sophisticated, architecture-specific optimizations that naturally make code much less portable. However, a compiler framework based on a multi-level IR can seamlessly apply both architecture-specific and architecture-independent optimizations, achieve optimal performance, and still maintain the high-level language specifications and semantics.

\begin{table}[t]
\small
\centering
    \caption{Performance speedup of re-ordering multi-operand expressions.}
    \begin{tabular}{|l|r|}
        \hline
        \textbf{Multi-operand Tensor expressions} & \textbf{Perf.} \\ 
        \hline
        \hline
        A[c,d,m,n] * B[i,n,a,d] * C[m,c]           & 23.9\\
        A[d,c,m,n] * B[i,n,a,d] * C[m,c]           & 21.1 \\
        A[c,d,m,n] * B[i,d] * C[n,a] * D[m,c]      & 4.9 \\
        A[d,c,m,n] * B[i,d] * C[n,a] * D[m,c]      & 4.5 \\
        A[m,n,e,j] * B[e,i] * C[a,m] * D[b,n]      & 1.4 \\
        A[m,n,f,e] * B[e,i] * C[f,n] * D[a,b,m,j]  & 1.4\\
        A[m,n,e,f] * B[a,m] * C[f,n] * D[e,b,i,j]  & 2.2 \\
        A[m,n,e,f] * B[b,m] * C[f,n] * D[e,a,j,i]  & 1.8 \\
        A[n,m,e,f] * B[a,m] * C[f,n] * D[e,b,i,j]  & 2.2 \\
        A[n,m,e,f] * B[b,m] * C[f,n] * D[e,a,j,i]  & 1.8 \\
        A[m,n,e,f] * B[e,i] * C[f,n] * D[a,b,m,j]  & 1.5 \\
        A[m,n,e,f] * B[e,j] * C[f,n] * D[b,a,m,i]  & 1.7 \\
        A[m,n,f,e] * B[e,j] * C[f,n] * D[b,a,m,i]  & 1.5 \\
\hline
\end{tabular}
\label{table:mutli-op}
\end{table}

\begin{table}[]
  \small
    \centering
    \caption{Characteristics of the emulated GEMM hardware.}
    \begin{tabular}{|l|r|r|r|}
         \hline
        & \textbf{16x16} & \textbf{64x64} & \textbf{256x256} \\
         \hline
         \hline
         \textbf{Perf. (cyc)}    & 131     & 1026    & 32770       \\
         \textbf{Avg. Power (mW)}      & 5.077   & 13.639  & 73.7972     \\
         \textbf{Avg. Area (uM$^2$)}   & 55827   & 224068  & 4.097e+06 \\
         \hline
    \end{tabular}
    \label{tab:gemm}
\end{table}

\textbf{Multi-Operand Expression} Differently from a library-based approach, where programmers need to match the library-defined (binary) APIs, \name{} can analyze tensor expressions and optimize the order in which each operation is executed. As stated in Section~\ref{sec:opts}, although the final results of a multi-operand tensor contraction do not change despite of the execution order of the contractions, the number of operations does change, hence some ordering provides higher performance than others.
In particular, when using a library (which normally provides an interface for contracting two tensors), programmers either have to figure out the optimal ordering manually a \emph{priori} or may incur in performance loss if they follow the natural order of the tensor contractions. \name{}, instead, automatically analyzes the entire expression and breaks it into binary operations properly ordered to achieve the highest performance by minimizing the number of overall operations. 
We evaluated 118 tensor contraction expressions from two NWChem methods that involve 3 and 4 operands.
Table~\ref{table:mutli-op} shows that the multi-operand optimization reduces the total number of operations and provides performance speedup, up to $23.9\times$. 
We do not report the other expressions, as the natural order coincides with the optimal ordering.
Note that in these experiments, we have employed the same optimizations introduced in Section~\ref{sec:opts}, only difference is that the baseline executes contractions in the natural order of the expression whereas the multi-operand expression optimization identifies an ordering that reduces the overall number of operations. 

\textbf{Case Study: Designing Custom GEMM Accelerators} Our final experiments use \name{} to perform co-design for a custom GEMM data-parallel accelerator. The main idea is to identify the best accelerator to perform GEMM operation in the TTGT method to solve tensor contractions. In this case, the GEMM accelerator is considered a ``hard accelerator'', as opposed to the ``soft accelerator'' for x86 employed in the previous section. 
We leverage \name{} modeling capabilities and combine the code generated by our compiler framework with the timing estimates produced by Aladdin models of the GEMM designs. In particular, we replace the x86 micro-kernel with a delay that represents the execution time of the innermost GEMM computation on the hardware accelerator.
We analyze three possible scenarios: small ($16\times16$), medium ($64\times64$), and large ($256\times256$). Table~\ref{tab:gemm} reports the hardware characteristics of the three designs 
in terms of performance, area, and average power. For comparison, consider that an Intel Ivy Bridge measures 160 $mm^2$ while an NVIDIA Volta GPU die measures 815 $mm^2$, which are 2,867x and 14,606x bigger than the 16x16  accelerator. 

Figure~\ref{fig:comparison} also reports the performance of a system that features custom GEMM hardware accelerators (dashed lines). The plot shows that hardware accelerators may substantially increase performance for compute-bound tensor contractions, such as the last 11 contractions, and achieve up to 156 GFLOPS, $3.1\times$ speedup over the same code employing a ``soft'' AVX512 accelerator. 
By comparing the results in Figure~\ref{fig:breakdown} and~\ref{fig:comparison}, it is evident that the contractions that most benefit from tiling (and thus become compute-bound) also greatly benefit from hardware acceleration. 
The plot also shows an important point: while it seems intuitive that larger accelerators provide higher performance, this is not always the case in our experiments. 
There may be several reasons for this behavior, including large carry-over loops, 
computing GEMM for non-square matrices, caches that are not large enough to contain all the data, etc. 
Figure~\ref{fig:comparison} does show that tensor contractions that are compute-bound with smaller hardware accelerators become memory-bound with the largest GEMM design. We infer that the lowest-level cache does not have sufficient storage to feed such large accelerators or to support data reuse. 
The actual point of co-design is, indeed, to figure out those trade-offs and select the best accelerator for the particular workload ($64\times64$) instead of the best accelerator from the single operation ($256\times256$).
\section{Related Work}
\label{sec:relwork}

Among the compiler-based approaches for tensor algebra, the Tensor Contraction Engine (TCE)~\cite{hirata_tce} is an early effort as a compiler framework that automatically optimizes dense tensor contractions in the quantum chemistry.
TACO compiler is a C++ library that employs compiler-based techniques for dense and sparse tensors.
TACO enables automatic code generation for a wide variety of linear and tensor algebra operations while supporting different storage formats.
TACO provides similar notation to \name{} TA language to express tensor expression, although programmers need to invoke object methods to pack/unpack data structures. 
Unlike TACO,~\name leverages core compiler optimizations, such as tiling or loop ordering, 
and supports multiple back-ends via LLVM.



There has been a lot of work on library-based approaches. 
The FLAME~\cite{flame_02} focuses on formal description of linear algebra methods on matrices and the derivation of optimized implementations for distributed-memory systems. 
Later works~\cite{elemental_13,symmetric_14,summa_16} extend the framework for multi-dimensional tensor operations.
The Cyclops Tensor Framework (CTF)~\cite{solomonik2014massively} focuses on distributed computation of tensor operations.
\code{libtensor}~\cite{libtensor_13} focuses on describing block tensor operations using \code{C++} templates. 
Recent work, such as ITensor~\cite{fishman2020itensor} and TensorNetwork~\cite{google_tensornetwork}, employs tensor networks to represent contractions of several tensors. 
Libraries-based approaches are easy to use but force scientists to re-arrange algorithms and implementations around the library APIs. This implies that, among others, library-approaches rarely support arbitrary tensor expressions. Moreover, libraries are typically implemented for specific architectures and may require heavy modifications to run on different heterogeneous systems.
\name{}, on the other hand, provides a programming interface close to the Einstein mathematical notation, supports arbitrary and mixed tensor expressions, and support execution on different architecture via LLVM backends.

To the best of our knowledge, none of the tools and libraries available provide an easy path to perform co-design of hardware accelerators for tensor algebra computations. \name{} has been designed to 
support hardware accelerators and allows swapping optimizations 
in and out according to the target architecture.
\section{Conclusions}
\label{sec:concl}
The recent explosion of high-efficient and specialized architectures has dramatically decreased program portability and productivity. On one side, scientists prefer high-level, domain-specific languages that provide high-expressiveness and portability; on the other side, achieving high performance on modern architectures requires highly-tuned, architecture-specific implementations and support for custom hardware accelerators. 
This work presents \name{}, a novel compiler framework that supports tensor algebra operations, specifically those related to chemistry. \name{} consists of a high-level DSL, 
a multi-level IR, and a series of progressive lowering steps and program optimizations. 
\name{}'s multi-level IR approach allows us to change some of the dialects without the need to re-implement the entire IR stack and optimizations.
We show that the code automatically generated by \name{} outperforms hand-tuned tensor contractions that leverage state-of-the-art computational libraries across 30 tensor contractions from various domains. 
Our approach provides the distinct advantage of analyzing multi-operand expressions and identify the optimal ordering of tensor operations, 
achieving up to $23.9\times$ speedup over equivalent code that follows the natural order. 
Finally, we show that \name{} can also be used to perform co-design and identify the best GEMM accelerator for the tensor contractions under study. 
We plan to extend \name{} in various directions, including additional support for tensor algebra operations, support for sparse operations, and support for additional architectures. 
We also plan to open source \name{}.

\section{Acknowledgement}
This research is supported by PNNL Laboratory Directed Research and Development Program (LDRD), Data-Model Convergence Initiative, project DuoMO: A Compiler Infrastructure for Data-Model Convergence, and project Hybrid Advanced Workflows.

\bibliographystyle{ACM-Reference-Format}
\bibliography{refs}




\end{document}